\documentclass[letterpaper,10pt,twocolumn]{article}
\usepackage[pdftex]{graphicx}
\usepackage{usenix}

\usepackage{times,xcolor}
\usepackage{listings}
\usepackage{amsmath}
\usepackage{paralist}           %
\usepackage{balance}            %

\usepackage{enumitem}

\usepackage[scaled=0.82]{beramono}

   \usepackage[square,sort&compress, authoryear]{natbib}
\newenvironment{acks}{\section*{Acknowledgements}}{}

\usepackage{verbatim}           %
\usepackage{siunitx}
\DeclareSIUnit[]\sloc{SLOC}
\newcommand{\code}[1]{\texttt{#1}}

\usepackage{xspace}
\usepackage{authblk}
\usepackage[T1]{fontenc}

\usepackage[pdftex]{graphicx}
\setkeys{Gin}{keepaspectratio=true,clip=true,draft=false}
\graphicspath{{./imgs/}}

\usepackage[pdftex]{hyperref}
\hypersetup{nolinks=false}

  \newcommand{\name}{Pancake\xspace}
  \newcommand{\lionsos}{LionsOS\xspace}
  \newcommand{\lionsCite}{Heiser_VCJGNLDZAZWPDB_25}
  \newcommand{\creplang}{CrepLang\xspace}
  \newcommand{\looplang}{LoopLang\xspace}

\begin{document}

  \renewcommand{\sectionautorefname}{Section}
  \renewcommand{\subsectionautorefname}{Section}
  \renewcommand{\subsubsectionautorefname}{Section}
  \renewcommand{\appendixautorefname}{Appendix}
  \renewcommand{\Hfootnoteautorefname}{Footnote}
  \newcommand{\Htextbf}[1]{\textbf{\hyperpage{#1}}}

\renewcommand{\topfraction}{0.9}

\title{Verifying Device Drivers with \name}
  \author[1]{Junming Zhao}
  \author[1]{Miki Tanaka}
  \author[4,5,1]{Johannes {\AA}man Pohjola}
  \author[2,3]{Alessandro Legnani}
  \author[1]{\authorcr Tiana Tsang Ung}
  \author[1]{H.~Truong}
  \author[1]{Tsun Wang Sau}
  \author[1]{Thomas Sewell}
  \author[1]{Rob Sison}
  \author[3,4]{Hira Syeda}
  \author[4]{\authorcr Magnus Myreen}
  \author[6]{Michael Norrish}
  \author[1]{Gernot Heiser}
  \renewcommand\Affilfont{\footnotesize}
  \setlength{\affilsep}{0.5em}
  \affil[1]{UNSW Sydney, Australia \textit{(\{junming.zhao,miki.tanaka,thomas.sewell,r.sison,gernot\}@unsw.edu.au)}}
  \affil[2]{ETH Z\"urich, Switzerland \textit{(alegnani@ethz.ch)}}
  \affil[3]{University of Melbourne, Australia \textit{(hira.syeda@unimelb.edu.au)}}
  \affil[4]{Chalmers University of Technology, Sweden \textit{(myreen@chalmers.se)}}
  \affil[5]{University of Gothenburg, Sweden \textit{(johannes.aman.pohjola@gu.se)}}
  \affil[6]{Australian National University, Australia \textit{(Michael.Norrish@anu.edu.au)}}

\maketitle

\subsection*{Abstract}

Device driver bugs are the leading cause of OS compromises, and their
formal verification is therefore highly desirable. To the best of our
knowledge, no realistic and performant driver has been verified for a
non-trivial device. We propose \name, an imperative language for
systems programming that features a well-defined and
verification-friendly semantics. Leveraging the verified compiler backend of
the CakeML functional language, we develop a compiler for \name that
guarantees that the binary retains the semantics of the source
code. Using automatic translation of \name to the Viper SMT front-end,
we verify a performant driver for an Ethernet NIC.

\section{Introduction}\label{s:intro}

Device driver bugs are the leading cause of OS compromises, accounting
for the majority of the 1,057 CVEs reported for Linux in the period
2018--22 \citep{Linux:CVEs}---%
clearly they should be the \#1 targets of OS verification efforts.

While there have been prior efforts to verify
drivers~\citep{Alkassar_Hillebrand_08, Alkassar:phd, Duan_Regehr_10, Duan:phd,
Chen_WSLG_16, Kim_CKK_08, Penninckx_MSJP_12, More:msc},
to our knowledge none have yet succeeded on realistic, non-trivial
devices, nor have they presented any performance analysis of
the drivers verified.

Most of these efforts used highly manual interactive
theorem proving, and sometimes required drivers to be written and analysed
in assembly.
On the other hand, attempts to apply more usable methods like
model checking \citep{Kim_CKK_08} and automated deductive
verification \citep{Penninckx_MSJP_12} left significant gaps between
the analysed model and the real code.

Of the above, only \citet{Chen_WSLG_16} provided an end-to-end
verification story that preserved the driver's correctness
from a driver-appropriate systems programming language down to the
binary.
Likewise, with no formal semantics or verified compiler, recent proposals like
that of \citep{Chen_LZB_24} to verify drivers written in Rust
\citep{Klabnik:RPL} have no plan to close the semantic gap to the binary.

Devices are commodities: new ones are created all the time. Therefore, any practicable driver verification must have a high degree of
usability and automation.
Drivers are typically written in C,
but verification of C code is made needlessly expensive by C's complicated semantics.
In short, the situation demands a performant systems language with support for
usable automated verification, and a means of ensuring the properties of
verified drivers are preserved at the binary level.

In this paper, we present verification of a performant, real-world Ethernet
driver that demonstrates a new approach to
efficient development of performant and verified
device drivers, achieved through the choice of driver design and the use of
\name, a new imperative programming language
designed specifically for verification of low-level systems code.
Namely, we verify the \name version of an Ethernet driver for
LionsOS~\citep{\lionsCite}, which uses {\em single-threaded}, {\em
single-purpose} (``modular'') driver designs as advocated
by~\citet{Ryzhyk_CKH_09, Ryzhyk_ZH_10} and outperforms Linux
networking with its standard C driver.  The \name version of the
driver we verify also enjoys the excellent performance owing to the
simple design, with reasonable overhead coming from porting
to \name. It will be made available as a choice of network drivers
for \lionsos soon.

Our new programming language \name has several features that
facilitate verification, the most significant being:
\begin{itemize}[leftmargin=1.5em]
\item a \emph{verified compiler} from \name to binary that leverages the final
    stages of the verified CakeML compiler~\citep{Tan_MKFON_19}.
This guarantees that the semantics of drivers and other systems code written in
\name, will be preserved at the binary level.
\autoref{s:pancake} discusses
the overall structure of \name's semantics and the compiler.

\item an \emph{automated deductive verification} front-end for \name that
    leverages the Viper verification framework~\citep{Muller_SS_16}, a
    middle-end for various SMT solver-based verification back-ends.
This takes the form of (1) an annotation syntax for \name, and
(2) a transpiler from annotated \name to the Viper intermediate language (IL),
which we explain in \autoref{s:frontend}.
\end{itemize}
Using this support, we produce the first verification of a device
driver for a non-trivial device, a driver for a popular 1\,Gb/s Ethernet card
for the seL4-based \lionsos~\citep{\lionsCite}.
This paper reports on the design of
the \name language, mainly on its compiler and semantics, and on the
verification of the \name
Ethernet driver using the Viper front-end for \name.

Our experience shows that \name's automated deductive verification
support is easy to use for those with a systems development background,
allowing the verification of critical guarantees for practical drivers
(\autoref{s:verif}). The verification of our Ethernet driver took around
three person-months. The driver was ported from C to
\name and then verified by the one member of the team who had no prior
verification experience. We expect the process will be even faster once a
library of example verified drivers is available.
This demonstrates \name as a viable
alternative to C for systems-level code, but one with the advantage of
accessible end-to-end verification support.

Our evaluation in \autoref{s:perf} shows that the verified \name
driver performs very close to the C version.

In \autoref{s:discuss} we discuss the \emph{trusted computing base} (TCB) and threats to the
validity of \name's automated deductive verification story and end-to-end
semantic preservation, and our plan to address them with further verification.

\section{\name Language and Compiler}\label{s:pancake}

\subsection{The Rationale}

While C is the de-facto standard systems language, C's semantics has a
number of undesirable properties for verification: a
complicated memory model, underspecified evaluation order, and the
need to prove the absence of undefined behaviour at almost every step.
While the seL4 verification demonstrated that these challenges can be
overcome,
the cost was high:
\$\qty{350}{\per\sloc} just for verifying the C code \citep{Klein_EHACDEEKNSTW_09}, and this cost continues to impact
kernel evolution. Using a verified
compiler~\citep{Leroy_09} with the verification toolchain VST~\citep{Appel_11}
may help, but to date this has not resulted in verified real-world drivers.

Many attempts to make better systems programming languages
incorporate advanced language features to make safety
properties hold by construction.
For example, Cogent's linear type discipline prevents memory leaks~\citep{Amani_HCRCOBNLSTKMKH_16},
Rust's borrow checker enforces ownership and lifetimes~\citep{Klabnik:RPL},
and Cyclone incorporates garbage collection and ML-style polymorphism~\citep{Jim_MGHCW_02}.
Such features can eliminate whole classes of bugs, or at
least reduce bug density, but at the cost of complicating the language
semantics and implementation. Garbage
collection introduces unpredictable delays that are highly undesirable
in low-level systems code. Yet these approaches still fall
short of ensuring full functional correctness, and it is unclear how
helpful they are in achieving it.

Functional correctness proofs routinely rely on significantly
stronger properties than type systems typically guarantee. A stronger
type system can give more properties, but unless it is so
powerful (and undecidable) as to be a full-featured proof calculus,
functional correctness proofs will need more.

The information provided by a type system is only useful in verification if the type
    system is sound, but most practical languages have unverified type
    systems, or type systems with known soundness bugs.  Type systems
    can be verified~\citep{DBLP:journals/jar/NaraschewskiN99}, but
    type soundness proofs are delicate, and have subtle
    interactions with even minor language changes. Maintaining a type soundness proof for a living language
    can significantly bog down development.

Moreover, the safety guarantees of a language only hold if
    no backdoors are used.
    But low-level systems programs often need
    to break out of type-safe environments.
    For example, driver code must adhere to
    hardware-specified data locations, layouts and access protocols.
    Hence driver code in safe languages must use significant
    amounts of unsafe code, effectively escapes to
    C~\citep{Astrauskas_MPMS_20, Evans_CS_20},
    which mostly eliminates the benefits of safe languages.

Instead of adding more safety features to a language, which
    tends to complicate the semantics, we believe simple proofs require a simple formal semantics.
    Of course, the semantics must exist in the first place:
    despite years of research~\citep{Jung_JKD_18,DBLP:journals/corr/abs-1903-00982,DBLP:journals/corr/abs-1804-07608,DBLP:conf/tase/WangSZZZ18},
    there is still no complete formal specification of Rust.

We propose {\bf \name} as the solution---a radically minimal language that nonetheless
offers a sufficiently expressive interface for writing
low-level systems programs, such as device drivers, alongside a number
of advantages for formal verification.  Most importantly, the language
is completely specified by a straightforward formal semantics that
fits in a few hundred lines of HOL4 code, with a simple memory model,
no notion of undefined behaviour, and no ambiguities in evaluation
order.

\name is an unmanaged language with no static type system,
at a level of abstraction between C and assembly.
The data representation and memory models are kept as simple as possible.
The only kinds of data are machine words, code pointers, and structs.
Programs cannot inspect the stack, which simplifies semantics.
All memory is statically allocated; there is no equivalent of \texttt{malloc} and \texttt{free}.

\subsubsection{Non-Goals}

\paragraph{Concurrency}
User-level drivers running as separate processes do not need to be
multi-threaded.  Even for Linux in-kernel
drivers, \citet{Ryzhyk_CKH_09, Ryzhyk_ZH_10} demonstrate that
single-threaded drivers are feasible and performant. They furthermore
show that 19\% of Linux driver bugs are concurrency bugs, which are
automatically eliminated by such a design. Single-threaded drivers are
much easier to verify, and map well onto the modular design of
microkernel-based OSes. This approach is routinely used for drivers on
seL4, including in LionsOS, which outperforms Linux on networking
using this design \citep{\lionsCite}.

Therefore, \name has no concurrency primitives, and our
Ethernet driver has no internal concurrency. We will see that
this does not meaningfully inhibit performance.

\paragraph{Rust-level verification}

\name is not intended to replace Rust; rather, it targets a different niche.

Rust's safety features benefit developers by allowing them to
write more trustworthy code in the absence of verification, and
Rust-based verification tools like Verus and Prusti have been very
successful.
However, Rust is a complex language with no authoritative
formal semantics, putting any verification on soft
foundations. It also requires trusting a large, unverified compiler,
which has been found to yield memory-unsafe binaries from
typechecked memory-safe Rust code~\citep{speykious_cvers}.
Furthermore, ``unsafe'' code is required for
implementing device drivers in Rust, and existing tools have limited
support for verification of unsafe Rust.

This level of verification
may be sufficient for certain situations, but we aim for stronger
guarantees.  We take the compiler out of our TCB by fully verifying it,
based on \name's precise semantics. A simpler language with formal
semantics makes verification easier in both automated and
interactive settings. Overall, what we offer is a framework for
producing secure low-level components with a very small TCB,
through \name's formal semantics and verified compiler.

\begin{lstlisting}[gobble=2,firstline=2,float=th,tabsize=2,
  label={l:example},
  caption={\name code snippet (concrete syntax) with annotations.}]

  fun example(1 N, 1 shared_addr) {
    /@ requires mem_access(N) @/
    /@ ensures mem_access(N) @/
    /@ ensures N == old(N) @/
    /@ ensures shared_addr == old(shared_addr) @/
    var x = 0;
    var i = 0;
    while (true) {
      /@ invariant 0 <= i && i < N && x <= 42 @/
      /@ invariant mem_access(N) @/
      var new_x = lds 1 @base + @biw * i;
      if (i + 1 == N) {break;}
      if (new_x > 42) {break;}
      x = new_x;
      i = i + 1;
    }
    /@ assert x <= 42 @/
    !stw shared_addr, x;
    return 0;
  }
\end{lstlisting}

\subsection{The Language and its Semantics}\label{s:semantics}

\name looks and feels like a
traditional imperative language (see \autoref{l:example},
ignore the \verb|/@...@/| annotations for
now).  \autoref{fig:pancake-syntax} shows the current abstract syntax
of \name. The nuts and bolts should be
familiar to any programmer. Mutable
variables, \lstinline{if}, \lstinline{while}---nothing fancy. This is
a deliberate design decision: we want \name to feel simple and
familiar to systems programmers.
In our experience, this has been borne out in
practice: systems programmers familiar with C have found \name
easy to learn.

{
\begin{figure}[t]
\hspace*{-2mm}
  \[\footnotesize
  \begin{array}{rcl}
  \mathit{exp} & := & \mathsf{Const}\ \mathit{word}
                      \quad | \quad\mathsf{Var}\ \mathit{string}
                      \quad | \quad\mathsf{Label}\ \mathit{string} \\
               & |  & \mathsf{Struct}\ \mathit{exp}^\star
                      \quad | \quad\mathsf{Field}\ \mathit{num}\ \mathit{exp} \\
               & |  & \mathsf{Load}\ \mathit{shape}\ \mathit{exp}
                      \quad | \quad\mathsf{LoadByte}\ \mathit{exp} \\
               & |  & \mathsf{Op}\ \mathit{binop}\ \mathit{exp}^\star
                      \quad | \quad\mathsf{Cmp}\ \mathit{cmp}\ \mathit{exp}\ \mathit{exp} \\
               & |  & \mathsf{Shift}\ \mathit{shift}\ \mathit{exp}\ \mathit{num}
                      \quad | \quad\mathsf{BaseAddr} \\
               & |  & \mathsf{BytesInWord} \\
               & & \\
  \mathit{prog} & := & \mathsf{Skip}
                      \quad | \quad\mathsf{Dec}\ \mathit{string}\ \mathit{exp}\ \mathit{prog} \\
               & |  & \mathsf{Assign}\ \mathit{string}\ \mathit{exp}
                      \quad | \quad\mathsf{Store}\ \mathit{exp}\ \mathit{exp} \\
               & |  & \mathsf{StoreByte}\ \mathit{exp}\ \mathit{exp}
                      \quad | \quad\mathsf{Seq}\ \mathit{prog}\ \mathit{prog} \\
               & |  & \mathsf{If}\ \mathit{exp}\ \mathit{prog}\ \mathit{prog}
                      \quad | \quad\mathsf{While}\ \mathit{exp}\ \mathit{prog} \\
               & |  & \mathsf{Break}
                      \quad | \quad\mathsf{Continue}
                      \quad | \quad\mathsf{Call}\ \mathit{ret}\ \mathit{exp}\ \mathit{exp}^\star \\
               & |  & \mathsf{Raise}\ \mathit{string}\ \mathit{exp}
                      \quad | \quad\mathsf{Return}\ \mathit{exp}
                      \quad | \quad\mathsf{Tick} \\
               & |  & \mathsf{ShMemStore}\ \mathit{opsize}\ \mathit{exp}\ \mathit{exp} \\
               & |  & \mathsf{ShMemLoad}\ \mathit{opsize}\ \mathit{string}\ \mathit{exp} \\
               & |  & \mathsf{DecCall}\ \mathit{string}\ \mathit{shape}\ \mathit{exp}\ \mathit{exp}^\star\ \mathit{prog} \\
               & |  & \mathsf{ExtCall}\ \mathit{string}\ \mathit{exp}\ \mathit{exp}\ \mathit{exp}\ \mathit{exp} \\
               & |  & \mathsf{Annot}\ \mathit{string}\ \mathit{string}
  \end{array}
  \]
  \vspace{-1.5em}
  \caption{\label{fig:pancake-syntax}Abstract syntax of \name.}
\end{figure}
}

Another key design
concern is to give programmers direct access to low-level details
without the language getting in their way. This is part of the motivation for
\name's perhaps most radical design decision: no static type system,
and no distinction between different kinds of data. In sloganeering
terms, \name is a language where everything is a
machine word. For example, there is no distinction between pointers
and integers: it's all words.

We can treat a word as an
integer by adding to it, or treat it as a
pointer by dereferencing it. The programmer can
do pointer arithmetic freely, which is of course unsafe in
general. We do not attempt to make it safe; rather, we give it a
simple and well-defined semantics that can support formal
verification, without the need for complicated rules about (say)
pointer provenance. In this way, the historically minded reader may
find the language design closer to BCPL than C.

With this in mind, the data representation and memory models of \name are kept as simple as
possible.
There are only three kinds of data:
\emph{machine words},
\emph{code pointers}, and
\emph{structs} (whose fields are machine words, code pointers, or
nested structs).
Local variables are stack-allocated, and we do not allow
pointers into the stack.  Global data may be stored in a statically allocated  global memory region.

The operational semantics of \name is specified using \emph{functional big-step semantics}~\citep{Owens_MKT_16}.
In this style, the core of the semantics is a logical function that looks like an interpreter, but is not
necessarily executable.
This simplifies formal proofs of compiler correctness by making the semantics more amenable to term rewriting.
The semantics of a program is defined in terms of how it communicates with the outside world;
specifically, as a possibly infinite trace of I/O events, each of which %
denotes either a foreign function call %
or a shared memory load/store operation. The semantics is parameterised
on a  model of the outside world and the kind of I/O events that can
occur.

\subsection{Verified Compiler}
\label{s:verified-compiler}

\begin{figure}[t]%
    \centering
    \hspace*{-8mm}
    \includegraphics[height=0.68\textheight]{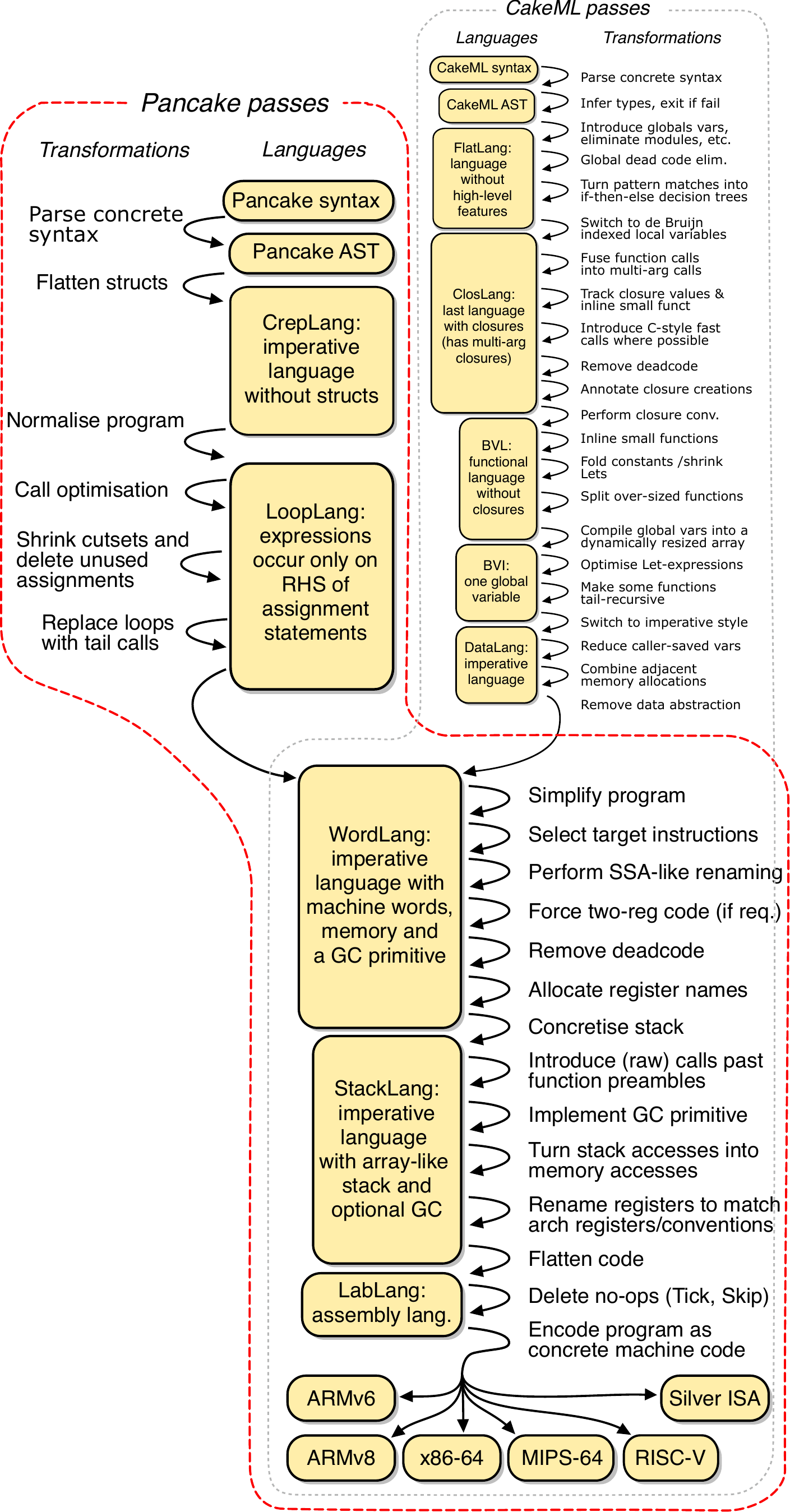}
    \caption{\label{fig:compiler}CakeML and \name compiler stack.}
\end{figure}

The \name compiler is verified, and thus not in the TCB. This
sidesteps the need for fragile validation of compiler output on a
program-by-program basis~\citep{Sewell_MK_13}.

We reuse the lower parts of the verified CakeML
compiler~\citep{Kumar_MNO_14,Tan_MKFON_19}, which compiles CakeML
programs to machine code for six architectures.
\autoref{fig:compiler} shows the CakeML compiler on
the right and the \name compiler on the left.  Arrows denote
compilation passes, and yellow boxes denote intermediate
representations.

CakeML itself is a high-level functional programming language,
unsuited for low-level systems programming.
For example, CakeML's memory management is all handled by the language runtime, and memory
allocation may trigger a stop-the-world garbage collection at any time.

By contrast, \name is explicitly designed to be unmanaged and
close to hardware, and to have no runtime.
Yet by integration into the CakeML ecosystem, it
can reuse many of the existing correctness proofs for the CakeML
compiler. The CakeML compiler provides backends for various target
architectures~\citep{Tan_MKFON_19} verified against detailed models
of the target ISA semantics.

The first few phases of the \name compiler go through two
new intermediate languages (ILs).
The first compiler phase flattens structs and
converts the programs to  \creplang, which is a stepping stone
into \looplang.  In \looplang, we compute minimal live sets and
divide loops (including their break and continue statements) into
tail-recursive functions that better fit the CakeML IL called
WordLang.  When the program under compilation is translated from
\looplang to WordLang, all loops are replaced with fast tail-calls,
as WordLang has no loops. Later, as in CakeML,
these tail calls will be realised as fast simple
jumps in the generated machine code.

Once we have entered WordLang, we use the CakeML compiler's
compilation phases, including its phases for instruction selection,
register allocation, concretisation of the stack, and, at the end,
encoding of the entire program into concrete machine code.

The compiler correctness proofs allow safety and liveness
properties of \name programs to carry over to the machine code that
runs them.

Our \name compiler has a special feature which produces
not only the output machine code, but also predicts the concrete maximum
stack size the program can use, as long as the input
program has recursive calls only in tail position.
We have proved that this bound suffices for running the \name program without
running out of stack space.  This allows \name to guarantee the absence of
premature termination arising from stack overflows.

The functional correctness proof of the \name compiler follows the
style of the CakeML proofs.  The top-level
correctness theorem states that the source \name program and the
compiler-generated machine code exhibit exactly the same observable
I/O events, if the compiler-generated code is run with enough stack space.

The compiler and the associated proofs are open source.%
\footnote{\url{http://code.cakeml.org} for source, or \url{http://cakeml.org} for pre-packaged versions. Note that \name is fully integrated into the CakeML compiler,
          rather than a stand-alone release.}

\subsection{Challenges with Device Drivers}\label{s:challenges}

Using CakeML as \name's backend compiler presents a number of issues when compiling device driver code.
First, CakeML assumes exclusive ownership of its statically allocated
memory region, and that this memory is not observable by the environment.
These assumptions
are not valid for memory-mapped device registers, which are special
memory locations used for interfacing with device hardware, nor do
they hold when devices directly write to memory (DMA).
They also do not fit the LionsOS model, which heavily depends on
shared-memory communication between OS components \citep{\lionsCite}.

With this limitation, any interaction with the device from CakeML
would have to be
mediated by FFI calls to C, adding a layer of indirection to
the TCB. We eliminate this indirection in \name by adding native support
for shared memory.

A key challenge here is that the ISA models used by the
compiler proofs~\citep{Tan_MKFON_19} are inherently sequential, and
\qty{>100}{\kilo\sloc} of proofs inextricably
rely on this fact.
While previous work has integrated driver models for
specific devices directly into an ARM ISA model with an interleaving
semantics~\citep{Alkassar_HKRT_07}, this approach would require us to
abandon much of the existing proof base, and would not provide the
modularity needed to target multiple ISAs and devices.  We
therefore parameterise the language semantics on a model for shared
memory, which supports proof reuse and provides flexibility for
incorporating arbitrary devices.

We add shared memory semantics to all compiler ILs from \name
downwards, and verify the full compilation chain.
We give load/store operations on shared memory regions a special
syntax (\code{ShMemLoad} and \code{ShMemStore}
in \autoref{fig:pancake-syntax}). In the semantics, these
operations are considered observable events. At the end of the
compiler pipeline they are converted to ordinary loads and stores.

Another challenge arising from the CakeML lineage is the issue of function entry points.
The CakeML compiler expects code to only ever be called via
the \code{main} function, and to exit entirely when \code{main}
returns. Moreover, \code{main} is preceded by lengthy initialisation,
and does not support parameters.

Device drivers in \lionsos must support several functions as entry
points, and for these to support re-entry.  Thus, using CakeML's
execution flow demands workarounds, such as branching in the main
function based on data indirectly passed through memory, and manually
editing the compiler output to return to the caller. Even then, a
substantial performance penalty arises from repeated
re-initialisation. This is clearly neither scalable nor performant.

We address this by implementing support for multiple entry points.
A function declaration can be tagged with the \code{export} keyword,
which extends the generated compiler output to
expose the function to the calling conventions of the target platform, and to restore the
state initialised by the main function when the exported function is called.
This circumvents the re-initialisation, and handles argument passing and
returning to the caller without programmer intervention. It enables
calling into the \name driver as if it were written in C.
The top-level compiler correctness proof currently accounts only for
the \texttt{main} entry point in that some of the assumptions are too
specific to initialisation, which need to be relaxed to account for
the re-entry points.

\subsection{Verification Approaches}
For the purpose of using \name to produce high-assurance device drivers,
the verified compiler (\autoref{s:verified-compiler}) is only half of the story: we must also be sure that the \name code implements the device driver correctly (incorrect source $+$ verified compiler $=$ incorrect machine code).

The gold-standard (but potentially expensive) way to verify source code is to use
interactive theorem-proving~(ITP), as done by (for example)
the seL4~\citep{Klein_AEMSKH_14} and CertiKOS~\citep{Gu_SCWKSC_16}
kernels, and the CompCert~\citep{Leroy_09}, and CakeML~\citep{Kumar_MNO_14} (and
thus \name) compilers. As
\name{} is an imperative language, it would be natural to use an
ITP-based implementation of Hoare logic.
We describe preliminary and future work on ITP-based approaches
to source verification in \autoref{s:itp-for-pancake}.

For productivity, given the commodity nature of devices, we
use automated deductive verification, rather than an ITP. Using
Viper~\citep{Muller_SS_16} (and its realisation of Hoare logic)
increases the size of the trusted computing base (see
\autoref{s:validity}), but can be expected to be less expensive in terms of
developer time and expertise. Our case study bears this
out.

To support this, \name{} syntax includes Viper annotations (visible
in~\autoref{l:example}). They resemble comments in the
source code, and can appear at the top level and within function
bodies. Top-level annotations are used, for instance, to specify
function contracts, while in-function annotations can be used for
loop invariants. The \name compiler ignores annotations, but
the Viper transpiler (\autoref{s:frontend}) preserves the annotations so
that verification can be carried out.

\section{A Viper Front-end for \name}\label{s:frontend}

Our verifier transpiles annotated \name code into the Viper IL,
then verifies the generated Viper code
using Viper's symbolic execution backend
Silicon~\citep{schwerhoff2016advancing}.
This approach is comparable to that of Viper front-ends such as
Gobra~\citep{wolf2021gobra} for Go and
Prusti~\citep{astrauskas2022prusti} for Rust.

Our transpiler first uses a diagnostic feature of the \name compiler
to extract the abstract
syntax tree (AST) of the input program, which includes \name-level
(1) annotations of the kind shown in \autoref{l:example} as well as
(2) hooks into a device model that we will describe in
\autoref{s:viper_model}.
It then translates that AST into our chosen encodings for \name's
variables, values, memory locations, and annotated logical assertions over
these, as expressed in Viper IL code.

The rest of this section documents and explains the most interesting of
these encoding decisions, those concerning our encoding of \name's machine word
type, and our encoding of (shared and unshared) memory in Viper.
Our objective an encoding that maximises the performance of the
resulting queries to Viper's backend, while still being sound---that is, it
should never produce a Viper query that is verifiable as true, when the
statement about \name code that it is supposed to encode is false.

We aim for most produced queries to take seconds or minutes to verify,
to make verification usable as an active part of a driver developer's workflow.
Our overriding concern is to prevent the transpiler from producing queries
that cause the Viper SMT-based back-end either to diverge or to take a
prohibitively long time to be used for continuous integration testing.
We leave proving the soundness of the encoding to future transpiler
verification work (see \autoref{s:discuss}).

\subsection{Machine-Word Encoding}

For the best query performance, we encode \name's machine-size word variables
as integers in Viper.

As a language whose only primitive type is machine-size words,
\name's word variables have bitvector semantics and
overflows are well-defined behaviour, i.e., the variable is
wrapped modulo the word size.

To preserve this semantics when encoding these as integers in Viper, which are signed and unbounded, one option is to
treat arithmetic operations as modulo the word size. However, this approach significantly
slows verification due to the mixing of arithmetic that is linear (such as
addition, subtraction etc.) versus non-linear (such as modulo and bitvector
operations) with respect to Viper's unbounded signed integer space.
Poor query performance due to mixing of different theories that require
handling by different solver strategies is a well-known hazard for SMT solvers
\citep{Jovanovic_Barrett_13}.

Instead, since overflows are rarely intended behaviour,
we adopt Prusti's approach, treating
overflows as verification failures. We do this by checking the bounds of every variable
after having unrolled all arithmetic operations into three-address code.
We make this decision to ensure soundness at the mild expense of disallowing
intended machine word overflows. These were used in the original C implementation
of the driver but could be avoided with a one line change in the \name version.

Another concern is that bitvector operations are also non-linear with respect
to Viper integers, causing similar performance concerns.
Observing that almost all bitvector operations in our driver occur as part of
bit masking, shifting etc.,~for accesses to device memory as part of a well
defined device interface, we instead abstract this device interface, which we
cover in \autoref{s:viper_model}. Additionally we precompute constant expressions
and apply heuristics to rewrite common bitvector operations, e.g.\ \texttt{x\&255}
is rewritten as \texttt{x\%256}. Although the rewritten operations still use
non-linear arithmetic, performance is improved compared to bitvector operations.
This eliminates the vast majority of bitvector operations, with the remaining
few in our Ethernet driver turning out to be reasonably performant as Viper queries.

\subsection{Local Memory Encoding}

As mentioned in \autoref{s:pancake}, \name disallows pointers to stack
variables, which simplifies modelling them in Viper: we do not have
have to manage any kind of access permission or to check for
references to invalid memory.

For all other memory,
the memory model of \name poses unique challenges due to its assembly-like nature, such
as the lack of first-order support for arrays and reliance on pointer arithmetic for
memory operations. Due to the lack of information about what different memory regions
represent---be they arrays, structs or other data structures---we must adopt
the most naive modelling approach: with the exception of shared and device memory
(described in the next section),
we encode memory as an array of words.
This approach results in non-idiomatic Viper code but captures \name's
native word-size treatment of
memory operations accurately.

The various memory regions a function needs access to, and with which permission, are added via
annotations encoded as an iterated separating
conjunction~\citep{muller2016automatic}, which can be verified efficiently in Viper.

\subsection{Shared and Device Memory Encoding}\label{s:viper_model}

As mentioned in \autoref{s:challenges},
\name includes dedicated load/store operations for shared memory regions,
whose behaviour in a sense resembles
\texttt{volatile} variables in C, whose accesses cannot be reordered or optimized away by the compiler.
Thus, driver code uses these primitives for accesses to device memory, as well
as accesses to memory shared between it and other OS components.
Given this sharing, we cannot encode shared memory in Viper
simply as an array of words, because we cannot rely on its contents not to
change between accesses.

Device register accesses, needed for implementing drivers, make extensive
use of bitwise operations to access the correct bits. As stated before, these
operations result in bad performance so we seek to avoid them where possible.

\begin{lstlisting}[gobble=2,firstline=2,float=th,tabsize=2,
  label={l:shared-annot},
  caption={Method signature of shared memory store in device model and the
	corresponding top-level annotation in \name, specifying that shared memory
    accesses to the address range \texttt{lower..upper} should transpile to a
    Viper invocation of \texttt{store\_rx\_free} or \texttt{load\_rx\_free}.}]

  method store_rx_free(heap: IArray,
      device: Ref, addr: Int, value: Int)

  /@ shared rw u64 rx_free[lower..upper] @/
\end{lstlisting}

To limit these, we model accesses to shared memory as separate Viper method
calls, which the driver developer should specify in an external Viper file
representing the \emph{device model} for the driver's target device.
These methods define valid operations for specific address ranges corresponding
to particular device registers and memory regions. They also specify \texttt{requires} and
\texttt{ensures} clauses for the hardware interfaces, as assertions in terms of a global
device state and the non-device memory.
This is a good fit with how the \name semantics
models shared memory operations (\autoref{s:pancake}) as observable events,
whose interpretation is parameterised on a model of the environment---%
a semantics which is also preserved by the \name compiler.

For interactions via shared memory with other OS components,
we make it the responsibility of the driver developer to specify a
\emph{neighbouring component model} similarly in a separate Viper file.
The developer should use this to capture the guarantees the driver should meet
for the Viper verification to enforce, as well as any assumptions about the
behaviour of those neighbouring components with respect to the shared memory.

We then provide a syntax for top-level \name annotations that allow the
driver developer to specify the correct method(s) to use for a shared memory
operation, according to the address range the driver interacts with, allowing
the transpiler to infer automatically which Viper method the produced Viper
model should invoke in place of the shared memory interaction (see \autoref{l:shared-annot}).

This way, the driver developer is not required to specify the model
of the device or neighbouring component before they implement their driver in
\name.
Furthermore, they do not have to modify their \name driver's shared
memory accesses after specifying these models---they instead add extra
annotations to specify which Viper methods the transpiler should produce in
place of loads/stores to given shared memory addresses.
This approach ensures the separation of driver implementation and device
specification, whilst also improving verification speeds.

\section{Verified Ethernet Driver}\label{s:verif}

\subsection{Overview}

We formally verify a single-core i.MX Ethernet driver for \lionsos \citep{\lionsCite}.
\lionsos uses a simple OS-side interface for device drivers
with zero-copy shared-memory communication for network data.
The interface employs lock-free, bounded, single-producer,
single-consumer (SPSC) queues that contain meta data buffers for
data addresses and lengths,
as well as control information for signaling requests.
The driver synchronises with the
rest of the OS via semaphores (implemented as seL4 Notifications).

The target driver, implemented in \name, controls the MAC-NET
1\,Gb/s Ethernet core common to NXP i.MX 8M Mini, Dual, QuadLite and Quad
Applications Processors present in various Arm-based NXP system-on-chips (SoC),
operating as a network interface card (NIC).
The NIC uses DMA descriptor rings for passing the addresses of data packets.
This driver will be made available for \lionsos as a choice of network drivers alongside the C version.
The driver incorporates formal specifications through annotations using
Hoare logic and the Viper verification framework.

\begin{figure}[t]
  \centering
  \includegraphics[width=1.0\linewidth]{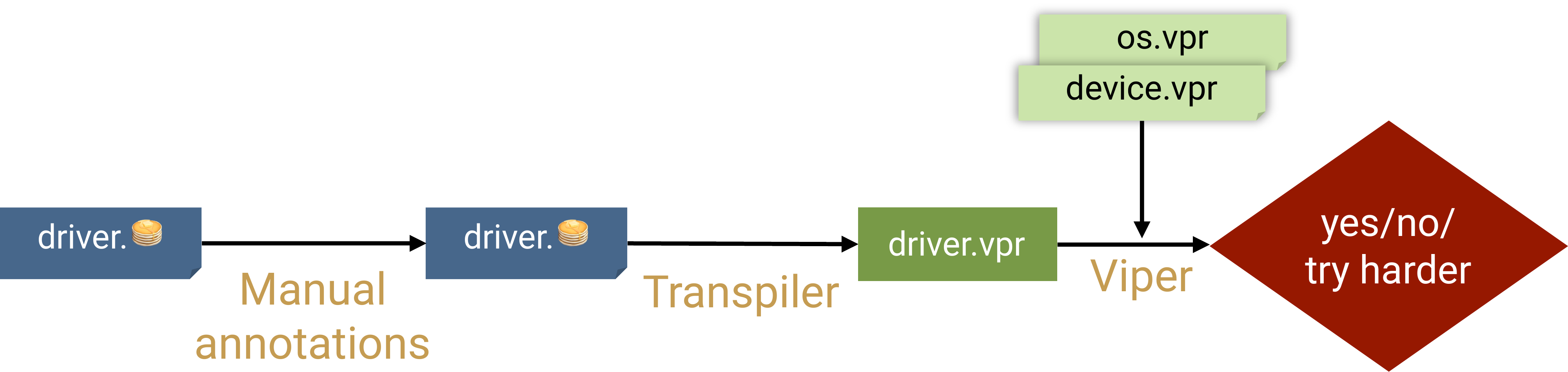}
  \caption{\label{fig:workflow}Driver verification workflow.}
\end{figure}

\begin{table}[h!]
  \small
  \centering
  \begin{tabular}{|| c | c ||}
    \hline
    \textbf{Component} & \textbf{Line Count} \\
    \hline\hline
    C Driver & 351 \\
    \hline
    \name Driver & 411 \\
    \hline
    \name annotations & 558 \\
    \hline
    Viper Device model & 391 \\
    \hline
    Viper OS interfacing model & 298 \\
    \hline
  \end{tabular}
  \vspace{0.5em}
  \caption{\label{tab:LoC}Lines of code comparisons between
  the Ethernet driver in \name implementation and C implementation,
  and the lines of code of our verification artifacts.}
  \vspace{-1.5em}
\end{table}

\paragraph{Verification workflow}

\autoref{fig:workflow} shows the resulting verification workflow. The
programmer annotates the \name source of the driver and processes the
resulting source with the transpiler, which produces the Viper input.
The programmer also supplies Viper specifications of the
device interface as well as the interface between the driver and the
rest of \lionsos. These are then processed by Viper, which returns a result (proved or
falsified) or times out.
\autoref{tab:LoC} shows the total size of our verification artifacts.

\paragraph{Device model}
We model the hardware NIC device and its non-deterministic state in
Viper specifications. The model is then connected to \name driver
source code after transpilation, using the top-level \name annotations
as described in \autoref{s:viper_model}.

The device state consists of RX (receive) and TX (transmit) hardware
descriptor rings, with each ring decomposed into three integer sequences
representing data addresses, lengths, and bitfields.
These integers represent machine words and are converted to bitvectors
when bit-level operations are required. As discussed above, by isolating the descriptor components,
such a decomposition simplifies verification and reduces SMT solver complexity,
particularly when using non-linear arithmetic and bitvector operations.
Each shared memory operation corresponds to a Viper method in the device model,
with memory loads returning non-deterministic values within valid ranges,
and memory stores satisfying both address and value requirements.

We model the non-determinism of the values the driver could possibly
obtain from the device by underspecifying its interface methods,
forcing the verification of the calling context of the method
(i.e.~the \name code that invoked the shared memory operation)
to account for a wide range of possible values
constrained only by their \texttt{ensures} clauses.
For example, \texttt{load\_EIR} in \autoref{l:method-snippet} ensures that
the return value is a valid unsigned 32-bit word value, as captured by
\texttt{bounded32(retval)}.
Verification must subsequently succeed for any value returned by that interface
that satisfies the \texttt{ensures}, i.e.,~any unsigned 32-bit word value.

\paragraph{What is verified}
This verification work establishes the following four classes of critical
guarantees:
\begin{enumerate}[leftmargin=*]
\item Compliance with NIC device interfaces (\autoref{s:device-model}):
we verify that:
\begin{itemize}[leftmargin=*]
\item device memory accesses are within the right address ranges; and
\item values written to the device obey the device's requirements.
\end{itemize}
\item Compliance with neighbouring OS component interfaces (\autoref{s:spsc-model}):
we verify that the driver adheres to the network queue signalling protocols, specifically:
\begin{itemize}[leftmargin=*]
\item for consuming meta data buffers from network queues, the driver requests
 the OS wake-up signals OS if hardware rings have vacancy,
\item for providing meta data buffers to the OS, the driver notifies the OS's semaphores
 if and only if (i) a signal was explicitly requested by the OS and (ii) the queue
 state has been changed by the driver; and
\item the driver clears signal requests after notifying the OS (to prevent double signalling)
\end{itemize}
\item Guaranteed data integrity across transfers (\autoref{s:data-integrity}):
we verify that there is no data loss in the driver by enforcing integrity
check on packet meta data (addresses and lengths) on all pathways of meta data
transfers in the driver (note that, with LionsOS~\citep{\lionsCite}, the driver
never needs to access or map the packet data itself, and hence never
violates the integrity or confidentiality of the packet data); and
\item Memory safety through region isolation and restricted access controls (\autoref{s:mem-access}):
 we verify that access to TX-related states and RX-related states are correctly
 restricted to the parts of the driver responsible for transmitting and
 receiving, respectively.%
\end{enumerate}

\subsection{Device Interface Compliance}\label{s:device-model}

First, we enforce that the driver only accesses the parts of device memory that
comprise the NIC's hardware interfaces for packet receipt and transmission---
namely, the device's RX (receive) and TX (transmit) hardware descriptor ring
regions and other essential registers.
To enforce this, recall from \autoref{s:viper_model} that our transpiler
supports and looks for top-level \name annotations that specify which addresses
correspond to valid device interfaces, as illustrated in
\autoref{l:shared-annot}.
Our verifier will reject any \name drivers that attempt to invoke a shared
memory load or store operation on an address that is not covered by any such
annotations.

\begin{lstlisting}[gobble=2,firstline=2,float=th,tabsize=2,
  label={l:method-snippet},
  caption={Examples of device register store/load interfaces in the device model
    as specified by Viper methods using \texttt{requires} and
    \texttt{ensures}, and corresponding top-level \name annotation.}]

  method store_EIR(device: Ref, addr: Int,
      value: Int)
    requires addr == (REG_BASE + EIR_OFFSET)
    requires value == IRQ_MASK
    requires valid_device(device)
    ensures valid_device(device)

  method load_EIR(device: Ref, addr: Int)
    returns (retval: Int)
    requires addr == (REG_BASE + EIR_OFFSET)
    requires valid_device(device)
    ensures bounded32(retval)
    ensures valid_device(device)

  /@ shared rw u32 EIR[REG_BASE + EIR_OFFSET] @/
\end{lstlisting}

Second, we verify that, whenever the driver interacts with these device
interfaces, it does so in the required way not to put the device into a bad
state, as specified by its documentation and captured by our device model.
To enforce this, we specify two kinds of \texttt{requires} clauses for device
interface methods, as illustrated in \autoref{l:method-snippet} for a
representative pair of examples, the \texttt{store} and \texttt{load} methods
for \texttt{EIR}, a particular device register:

\begin{enumerate}[leftmargin=1em]
\item
Method-specific requirements, such as the \texttt{store\_EIR} method's
\texttt{requires} of both an \emph{address} and a \emph{value} requirement---%
namely, that the address is the \texttt{EIR}'s and that the driver only ever
writes a particular \texttt{IRQ\_MASK} constant to it;
\item
Device-wide invariants---in this example, captured by \texttt{valid\_device}
in \texttt{store\_EIR}'s third \texttt{requires} clause.
For our NIC, \texttt{valid\_device} asserts that the state of the hardware
descriptors remains valid:
the bitfields are cleared and set properly according to the device's documented
specifications, data pointers are 32-bit width and byte-aligned, and data
lengths are within 16-bit bounds.
\end{enumerate}
The device state of our NIC device, as modelled by \texttt{device}
and asserted valid by \texttt{valid\_device(device)},
comprises the hardware descriptor rings,
with each ring decomposed into three integer sequences
representing data addresses, lengths, and bitfields.
These integers represent machine words and are converted by Viper to bitvectors
when bit-level operations are required.
Our decomposition of these descriptor components into native Viper integers
thus simplifies verification and reduces SMT solver complexity,
particularly when using non-linear arithmetic and bitvector operations.

Note that, while the presence of \texttt{valid\_device} in the
\texttt{requires} clauses of device interface methods
requires the driver not to violate the validity of the device state,
its presence also in the \texttt{ensures} clauses of all device methods
specifies that we can assume the device itself will maintain that same validity
invariant through all driver-device interactions.

\subsection{OS Communication Protocols}\label{s:spsc-model}
Using much the same approach just described in the previous section for
specifying valid device interactions,
we constrain the driver's access to shared memory
regions for network SPSC queues through annotations when interfacing with
\lionsos components. We also model the shared SPSC queues
non-deterministically to verify the driver maintains protocol compliance%
---assuming the neighbouring OS component maintains it too---%
without assuming specific state values.

\subsection{Data Integrity}\label{s:data-integrity}
To ensure reliable data transfer between the OS and the device during
the translation between hardware descriptor and SPSC formats, we verify that
the driver maintains data integrity by tracking packet addresses and lengths.

For example, we check that the given data address and data length are stored
properly after updating the TX hardware descriptor ring, as shown in \autoref{l:assert-snippet}.

\begin{lstlisting}[gobble=2,firstline=2,float=th,tabsize=2,
  label={l:assert-snippet},
  caption={Data integrity verification example in annotated \name.
    For brevity, unwrapping of predicates referred to by
    the assertions is omitted.}]

  buffer = net_dequeue(os_tx_avail);
  update_tx_hw_ring(hw_tail, buffer);
  /@ assert(device.hw_ring_tx[hw_tail].data_addr
        == buffer.data_addr) @/
  /@ assert(device.hw_ring_tx[hw_tail].data_len
        == buffer.data_len) @/
\end{lstlisting}

We establish this integrity check on all pathways of data
transfers in the driver. We also verify data transfer completeness
by ensuring that within the driver, the number of SPSC queue operations align with the number of hardware
descriptor ring state changes, so there is no data loss in the driver.

\subsection{Memory Access Control}\label{s:mem-access}
In addition to the memory access constraints described in
Sections~\ref{s:device-model} and \ref{s:spsc-model},
we also verify that only the parts of the driver responsible for packet
transmission access any TX-related descriptor rings and SPSC queue state,
and likewise that only its packet receipt paths access RX-related state.

We enforce this using Viper's native permissions to specify access
controls,
in this way providing formal verification of
memory safety and region isolation.
In effect, our driver verification applies a separation logic-like principle by
partitioning the driver's global memory into RX and TX
regions\ (which reflects their use in \lionsos \citep{\lionsCite}).

\section{Performance Evaluation}\label{s:perf}

We now examine how our verified \name driver compares to the original
C implementation.

\name driver compilation time is a matter of seconds, and
the verification of the driver in full takes around 20 minutes
on a typical laptop. When verified separately, the device model takes
10 minutes and the driver's functions each take around 1 minute.

Our evaluation platform is an AVnet MaaXBoard with an NXP i.MX8MQ SoC,
having four Arm Cortex A53 cores capable of a maximum of 1.5\,GHz. We
run our measurements at a fixed clock rate of 1\,GHz to avoid the
need for thermal management.%
\footnote{Note that running at a low clock rate will magnify software
  overheads, and as such presents a pessimistic performance.}
The board has
2\,GiB of RAM and the on-chip 1\,Gb/s NIC specified earlier
(in \autoref{s:verif}).

The evaluation system runs a networking client on \lionsos. The client
simply receives data packets from the NIC and echoes them back. We use
an external load generator that sends an adjustable load (requested
throughput) to the target system, and measures the amount of data received
back (received throughput) as well as the latency. On the evaluation
system we also measure CPU load.

\begin{figure}[t]%
  \centering
  \hspace*{-2mm}
  \includegraphics[width=1.01\linewidth]{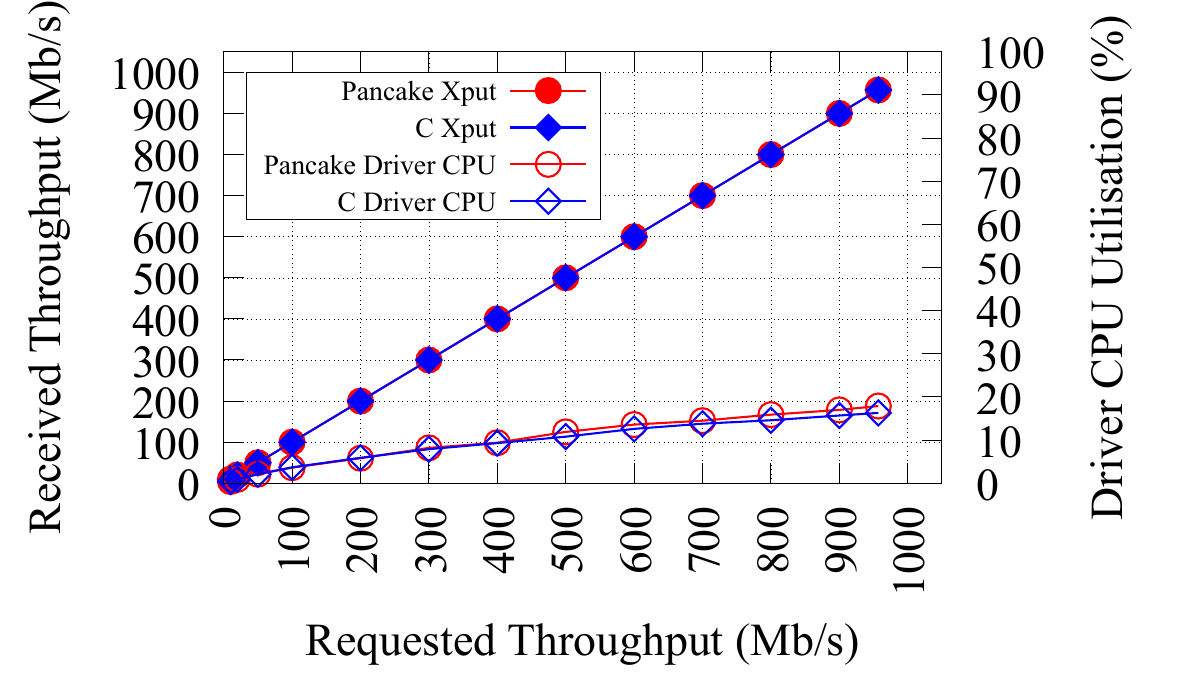}
  \caption{\label{fig:xput-cpu} Performance of Ethernet Driver written in \name vs C,
  in terms of achieved throughput (Xput) and Driver CPU utilisation.}
\end{figure}

\begin{figure}[t]%
  \centering
  \hspace*{-2mm}
  \includegraphics[width=1.01\linewidth]{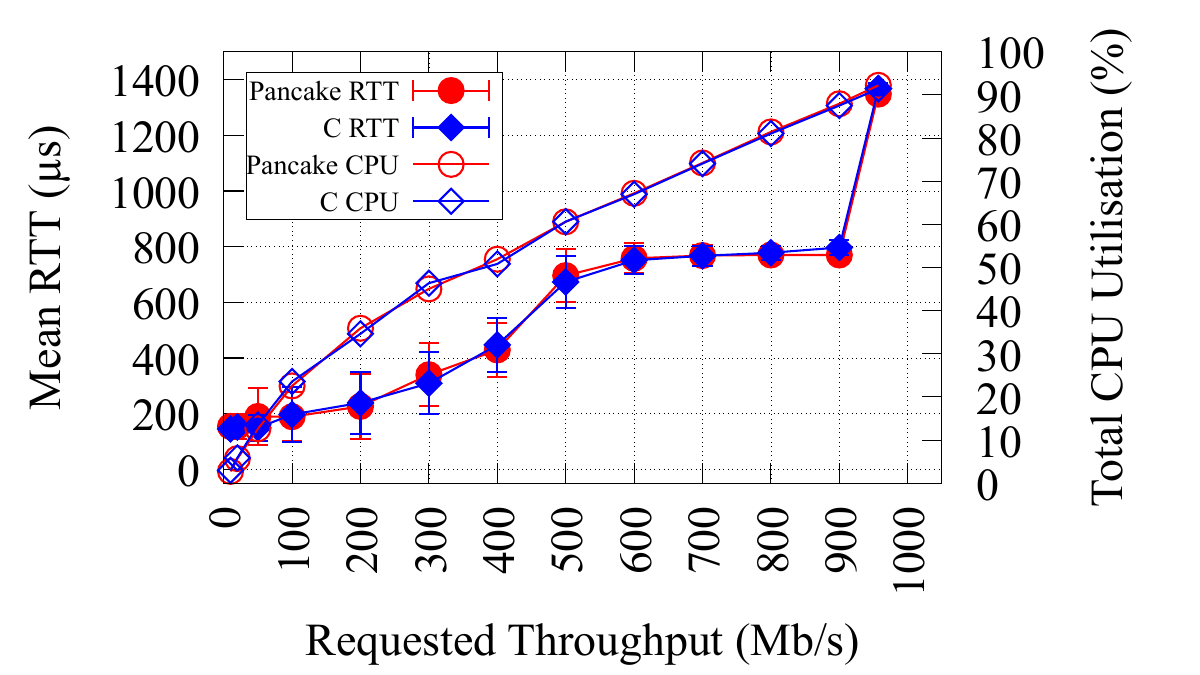}
  \caption{\label{fig:latency} Performance of Ethernet Driver written in \name vs C,
  in terms of average round trip time (RTT) with standard deviation and total CPU utilisation.}
\end{figure}

\autoref{fig:xput-cpu} shows the result. Like the C driver, the \name
driver has no problems
handling the requested load: the received throughput equals the
requested. The \name version of the driver uses slightly more CPU time
to handle the load than the C version, the difference is about
10\%. Viewed in the context of \lionsos using less than half the CPU
of Linux to handle a particular network load, the slightly increased
overhead of the \name driver is in the noise.

\autoref{fig:latency} puts this slight increase into context, by
looking at latency (RTT) and the \emph{overall} CPU use of the whole
system. The difference in CPU usage of the drivers becomes
unnoticeable. Similarly, the differences in latency is well within the
variance of the results. This is in the context of \lionsos using less than half the CPU
of Linux to handle a particular network load \citep{\lionsCite}.
We can summarise that the performance cost of
the verified driver is in the noise.

\section{Discussion}\label{s:discuss}

While our main results are formally verified, no proof about a
real-world artifact is ever fully complete and self-contained: there
will always be a \emph{trusted computing base} (TCB). The TCB, in
brief, is everything that we rely on for the correct operation of our
system, but which is currently outside the scope of the formal
verification effort. For a formal verification result, the TCB is the
main threat to validity.

In \autoref{s:validity}, we discuss our TCB. \autoref{s:future}
discusses alternative verification approaches we are pursuing that
would reduce the TCB further. Finally, \autoref{s:transpiler-usability}
discusses potential usability improvements to the verification front-end.

\subsection{Threats to Validity}\label{s:validity}

For the \name compiler correctness proofs, we trust: that the HOL4
theorem prover is a sound implementation of higher order logic; that
the official specification of the Arm ISA is correctly
implemented by the CPU \citep{Kanabar_FM_22}; and that an unverified
linker for connecting the \name binary to external code such
as \lionsos library routines is correct.
With these as the TCB, the \name compiler is verified, just
like the CakeML compilation passes that
it leverages, to preserve the
semantics of the \texttt{main} function of the source program.

Our verification results presented here establish certain observable properties
of the driver including its behaviour after reentry.
Putting these together,
the verified compiler guarantees that the binary obtained by compiling this driver should preserve
the verified observable properties, although we are not (yet) providing
links between the two results formally, in the sense that there is no theorem yet to state this
on a single, unified formalism.
Future work will strengthen
this by verifying the transpiler as well as extending the compiler proofs to
account for reentry points.

Moreover, while we plan to verify our transpiler from annotated \name
to Viper (\autoref{s:f-transpiler}), this is not done yet.
Until then, a mistake in the transpiler could produce a Viper query that the
underlying SMT-based backend can prove true, even if the property as specified
at the \name program point via Viper annotations is false.
Beyond this, we also trust the Viper verification infrastructure and SMT
solvers called by its backends to discharge the Hoare logic queries specified
using its input language soundly, i.e.,~only when true.

Like any device driver, we trust the device not to malfunction---that is, to guarantee
that it meets the \texttt{ensures} specifications we impose on return from
device interfaces in our device model.
This includes trusting that the device's initialisation process establishes a
valid initial state.
Verification of the device hardware and initialisation process
would be needed to gain further assurance of this.

Meanwhile, we assume that the neighbouring OS components comply with the
SPSC queue protocols, as mentioned in \autoref{s:spsc-model}, and that the
driver acts as the sole consumer or producer per queue.
We will need to leverage Viper's access permission system to establish
more sophisticated guarantees about concurrent accesses and thread safety properties.

Finally, we trust the operating system kernel and its user-level support
libraries not to crash or malfunction.
\lionsos runs on the seL4 OS microkernel, which is verified not to
crash~\citep{Klein_AEMSKH_14}, and relies on the seL4 Microkit support library,
whose main server loop has been verified not to exhibit undefined
behaviour~\citep{Paturel_SH_23}.
However, functional properties of individual seL4 system calls
relied on by Microkit have not yet been verified.

\subsection{Future Verification Efforts}\label{s:future}

We are continuing to implement more drivers for \lionsos in \name and
to apply the verification method presented in this paper to those
drivers firstly to produce more verified drivers but also to assess
and improve the verification process. We also have various
verification related future projects based on \name, including those we
discuss below.

\paragraph{Transpiler verification}\label{s:f-transpiler}

Proving the formal correctness of the transpiler implementation, as
mentioned in~\autoref{s:frontend}, is in progress.  This means that
the transpiler is currently included as part of the TCB, as discussed
in~\autoref{s:validity}.  What we aim to prove here is to establish
the validity at the level of the \name input of the statements proved
on at the level of the transpiled Viper IL code.
Such a proof would complete our end-to-end verification story, by allowing us to soundly infer the correctness of our initial \name code from
a successful verification run of the corresponding Viper IL code.
This still assumes the correct operation of the Viper toolchain which our verification step relies on.
Efforts exist to validate parts of this toolchain \citep{parthasarathy2024towards, gossi2016formal}.

We will do this proof using the theorem prover HOL4,
which \name compiler is implemented.  Though the current Rust
implementation of the transpiler contains 7 kLOC, a significant
portion of this code consists of workarounds for extracting the
underlying abstract syntax from an unparsed \name program without
interfacing directly with the existing \name parser. We can remove
these workarounds entirely by connecting the transpilation to the \name
parser and handling its formal syntax and the semantics directly,
which will simplify the implementation considerably, making the
verification easier.

\paragraph{Interactive theorem proving for \name programs}
\label{s:itp-for-pancake}

The formal semantics for \name and the compiler for it are developed
in the HOL4 interactive theorem prover. It would be natural to verify
a \name program entirely in HOL4. We have performed a number of small
experiments in this style, verifying simple example programs such as
a loop that sums the elements of an array. These experiments are an
important validation of the \name formal semantics. The compiler
correctness proof establishes that the formal semantics captures the
behaviour of the binary, but that is only useful if it is possible to
reason at the level of the semantics.

We have done some of these proofs by directly manipulating the
\name semantics in HOL4. We have also constructed a Hoare-style
precondition-postcondition logic for \name programs, which is a
standard approach to verification of programs in an imperative
language such as \name. This gives us a comparison point to our
Viper-based approach. Our Hoare logic is far more user-friendly than
directly working with the semantics, but far less user-friendly than
using the transpiler. One advantage of the Hoare logic is that its correctness
results compose directly with the compiler correctness theorem in HOL4,
producing a binary correctness theorem with a minimal TCB, useful for small
safety-critical programs whose correctness must be certified.

Another strength of HOL4 is its general logic, which allows us to phrase and
prove properties that we could not easily express in Viper. We are interested
in reasoning about concurrent scenarios. \lionsos encourages systems to be
composed from smaller components (such as our Ethernet driver), most of which
are internally single-threaded but execute in parallel. We think that studying
a parallel protocol in HOL4 and proving correctness of each \name component
against that protocol using Viper is an interesting direction for future work.

This composition of components will be correct if each component performs
the correct I/O actions. The I/O behaviour of a program can be captured in
a single value using interaction trees~\citep{Xia_ZHHMPZ_20}. These are a
new approach to program semantics that use an infinite coinductive tree
of input and output event nodes to model a program. We have developed an
alternative semantics for \name using interaction trees, and have experimented
with proofs that programs obey an I/O protocol using this semantics. This is
a promising approach to verifying, for instance, a whole network sub-system.

\subsection{Transpiler usability improvements}\label{s:transpiler-usability}

The current workflow involves transpiling \name into Viper for verification.
Whilst the transpiler allows for direct verification, error messages are tied
to the generated Viper code and are not reflected back to the \name code,
complicating debugging.
Additionally, while many annotations are automatically
inferred, some require manual specification, increasing the effort required from the
programmer.

In addition to the command line tool we have built an initial development environment for verified \name
based on Microsoft's VS Code framework.
As of yet this lacks some features like support for the separate Viper device model.
The transpiler supports verifying functions individually.
This could be integrated into the development environment to
allow for efficient re-verification
of only modified functions.

Despite these shortcomings, we expect that these tools can be combined and improved
to form a cohesive and usable toolchain for writing and verifying device drivers.

\section{Related Work}\label{s:related}

Among various verification efforts in the OS context, a number of prior
works have investigated verification of device drivers at
various programming language levels.
Within this space, researchers have employed different verification approaches
like model checking \citep{Kim_CKK_08} and interactive theorem proving
\citep{More:msc}.
Among these efforts, \citep{Penninckx_MSJP_12} developed their verification
using \mbox{VeriFast} \citep{Jacobs_SP_10}, a deductive verification approach
similar to ours. They also notably extended their analysis to include concurrency
properties beyond our current scope.
However, a common limitation across all these approaches was the significant gap
between the analysed model and the actual executable code---%
a gap which we narrow by using \name's verified
compilation stack.

Others, which we detail below, have done better at closing such gaps
(e.g.~in some cases providing support for direct access by drivers to
memory-mapped device ports, as we have), but failed to demonstrate scalability
beyond the simplest serial drivers -- among other reasons, by failing to
include any performance evaluation of the drivers they verified.

The earliest driver verification effort for a non-trivial device we are aware of
is that of \citet{Alkassar_Hillebrand_08, Alkassar:phd},
who verified a (still simplified) ATAPI hard disk device driver in
Isabelle/HOL interactive proof assistant \citep{Nipkow_PW:Isabelle}
as part of the Verisoft project.
Similar to our work, they verify their driver relative to a functional model of
the memory-mapped device -- in their case, based on a subset of the ATAPI
standard.
However, the type safety of the fragment of C they used for most of their OS
limited their ability to model direct access to memory-mapped device ports
directly from that language; consequently, they instead had to write and verify
their driver in a MIPS-like assembly language.

\citet{Duan_Regehr_10} presented a framework for verifying device drivers
integrated with the L3 model of ARM machine code
\citep{Fox_03, Fox_Myreen_10} for HOL4 \citep{Slind_Norrish_08},
with a UART driver as the case study.
Like \citet{Alkassar_Hillebrand_08} they did not support reasoning
about DMA, but \citet{Duan:phd} later added support in
the form of Hoare triples for device memory access scenarios.
The way we integrate shared memory access in
\name with the specification of \texttt{requires} and \texttt{ensures} as Viper
annotations is similar, and allows us to impose the requirements of our device model on our
driver's device memory access directly from the \name language.
\citet{Schwarz_Dam_14} further extend the L3 model to
support device drivers with DMA.  This goes further than the device
model of our paper, as the Ethernet device we verify is documented not
to interfere with the hardware ring indices, which are left under the
control of the driver.

As part of the CertiKOS project, \citet{Chen_WSLG_16} added support for
verifying drivers and integrating them with their OS verification
framework in the Rocq (formerly Coq) interactive proof assistant.
Unlike the above works, whose drivers were implemented in assembly,
the serial and interrupt controller device drivers \citet{Chen_WSLG_16}
verified are implemented in ClightX \citep{Gu_KRSWWZG_15}, an extension of the
CompCert Clight language \citep{Leroy_09} with extra instrumentation to support
\mbox{CertiKOS's} abstraction-guided approach to OS verification.
Like our work, their use of a verified compiler (a modification of CompCert) to
compile the driver down to binary gives some assurance that any properties
proved at the driver source level are preserved down to the binary.
However, driver verification in their framework requires interactive proofs in
Rocq, for a C variant whose proof relies on adding abstract state elements that
can influence program execution.
This is more disruptive to the original code than mere annotations or
typical ``ghost state'', and
arguably requires more formal
methods experience than automated deductive verification via annotations.

Unlike our work, none of the works above presented any analysis or
discussion their drivers' performance.

The Ironclad verification \citep{Hawblitzel_HLNPZZ_14} includes a
network driver.
They verify assembly code by translating it into Dafny
code along with assertions. They argue that verification at the level of
assembly avoids issues with the compiler; since we have a verified compiler
we prefer to verify the sources that the programmer writes. Like in our work, the
Ironclad authors annotate each driver function. They mostly establish memory
safety and address a public/private data distinction. There does not appear
to be any clear statement of the function of the driver or of functional correctness.
The Ironclad paper gives some performance data, showing overheads of up to two
orders of
magnitude for single requests to their verified versus their unverified system,
but no measurements of the driver alone or of the system under load, which would
be needed to make a direct comparison to our work.
They also note that performance is not their priority, whereas our goal
here is to achieve both of high guarantees and performance.

\citet{Erbsen_GCWC_21} offers an end-to-end story of a simple embedded system
that includes an Ethernet driver. However, this system is simple
enough that it does not require an operating system in the usual sense.
They also report ten-fold overhead as the performance results for the whole
system. Again, their work does not provide analysis on the driver by
itself, making a direct performance comparison difficult.

We are also aware of some current efforts by \citet{Chen_LZB_24}
to verify device drivers written in Rust \citep{Klabnik:RPL}
using the Verus automated deductive verifier \citep{Lattuada_HCBSZHPH_23}.
However, without a formal semantics let alone a verified compiler, the
possibility of end-to-end assurances for Rust-based drivers still seems remote.

Finally, there has been also been work on driver synthesis by
\citep{Ryzhyk_CKSH_09,Ryzhyk_WKLRSV_14} that took as input detailed
specifications of interfaces for (1) the device class the driver needs to
implement, the (2) device itself and (3) OS service it needs to provide to the
rest of the OS, written in a custom specification language.
Although, like in our work, their device interface included details such as
valid registers and their sizes, it also included more detailed elements like a
state transition diagram.
In our work, we have a model of device state that we use for specifying and
verifying the maintenance of invariants (the \texttt{valid\_device(device)}
assertion seen in \autoref{l:method-snippet} and explained in
\autoref{s:device-model})---this could in future form the basis for more
detailed, state machine-based specifications of internal device states.
Note, however, that this synthesis work could not deal with DMA.

\section{Conclusion}\label{s:concl}

This paper presents, to our knowledge, the first formal verification of a
demonstrably performant driver for a realistic, non-trivial device, the
Ethernet NIC common to a number of variants of NXP i.MX 8M processors.

It also introduces the \name systems programming language, designed especially
for systems-level code to be amenable to formal verification.
With \name, it makes two enabling contributions:
(1) a verified compiler that carries the semantics of \name down to binary,
leveraging CakeML's verified compiler backend; and
(2) an automated deductive verification front-end that takes \name with
Viper annotations, leveraging the Viper SMT-based verification framework.

This work shows that \name is usable for developing
verified, performant drivers.
A PhD student with a systems background and not much formal methods
experience was able to write and verify the aforementioned Ethernet driver in a
few person-months.
The \name driver shows performance very close to C.

This work paves the way for verified development of
performant device drivers---a leading source of OS vulnerabilities---as common-place
infrastructure.

\begin{acks}
  Many thanks to Alessandro Legnani's project co-supervisors Toby Murray, for his role
  in envisioning the use of Viper, and Peter M\"uller for his feedback on this.
  We thank also Krishnan Winter, Benjamin Nott, Craig McLaughlin and Remy
  Sasseau for their past and ongoing contributions to other aspects of Pancake.
  Finally, we thank Isitha Subasinghe and Zoltan Kocsis, for their help
  understanding SMT performance issues from theory mixing.

  The development of Pancake was made possible through the generous support of multiple
  organisations: the UAE Technology Innovation Institute (TII) under
  the \emph{Secure, High-Performance Device Virtualisation for seL4}
  project, DARPA under prime contract FA 8750-24-9-1000, UK's
  National Cyber Security Centre (NCSC) under project NSC-1686,
  and the German Agentur f\"{u}r Innovation in der
  Cybersicherheit GmbH (Cyberagentur).
  Finally, Junming Zhao gratefully acknowledges the topup scholarship donated by the winners of
  the 2023 ACM Software System Award.
\end{acks}

\balance
{\sloppy
  \footnotesize
      \bibliographystyle{plainnat}
    \bibliography{references}
}
\end{document}